\documentclass[11pt]{article}          
\usepackage[latin1]{inputenc}
\usepackage{amsmath}
\usepackage{amsthm}
\usepackage{amsfonts}
\usepackage{amssymb}
\usepackage{graphicx}
\usepackage{cleveref}
\usepackage{caption}
\usepackage{subcaption}
\usepackage{epstopdf}

\theoremstyle{definition}
\newtheorem{definition}{Definition}

\newtheorem{proposition}{Proposition}

\newtheorem{assumption}{Assumption}

\begin{document}
	%\title{Average Consensus-based Control over Wireless Fading Channels}
	\title{\LARGE \bf Exploiting the Superposition Property of Wireless Communication For Average Consensus Problems in Multi-Agent Systems}
	\author{
		Fabio Molinari\thanks{F. Molinari is with the Control Systems Group - Technische Universit\"at Berlin, Germany. {\tt\small molinari@control.tu-berlin.de}}
		\and
		S\l awomir Sta\'nczak	\thanks{S. Sta\'nczak is with the Network Information Theory Group - Technische Universit\"at Berlin \& Fraunhofer Heinrich Hertz Institute, Germany.{\tt\small slawomir.stanczak@hhi.fraunhofer.de}}
		\and
		J\"org Raisch\thanks{J. Raisch is with the Control Systems Group - Technische Universit\"at Berlin, Germany \& Max-Planck-Institut f\"ur Dynamik Komplexer Technischer Systeme, Germany. {\tt\small raisch@control.tu-berlin.de}
		\newline
		\newline
		{This work was funded by the German Research Foundation (DFG) within their priority programme SPP 1914 "Cyber-Physical 
			Networking (CPN)".}
		}
	}
	\maketitle
	\begin{abstract}
		This paper studies system stability and performance of multi-agent systems in the context of consensus problems over wireless multiple-access channels (MAC).
		We propose a consensus algorithm that exploits the broadcast property of the wireless channel. Therefore, the algorithm is expected to exhibit fast convergence and high efficiency in terms of the usage of scarce wireless resources. The designed algorithm shows robustness against variations in the channel and consensus is always reached. However the consensus value will be depending on these variations.
		%We design a consensus algorithm that stabilizes the system and is robust against variations in the wireless channel. Moreover simulations show that consensus is reached faster than in the case of conventional communication protocols such as TDMA. 
%		The cost for the performance gains is that the agreement value depends on the above mentioned coefficients. 
	\end{abstract}	
	\section{Introduction}
%	Consensus problems lie at the core of distributed computation, which has a long tradition in literature \cite{lynch1996distributed}. A typical consensus problem setup involves a multi-agent system in which the
%	%Distributed computation has a long tradition in literature \cite{lynch1996distributed} and its foundation is represented by consensus problems, 	where 
%	agents attempt to reach an agreement over a certain real-valued scalar or vector, which we call \textit{information state} \cite{degroot1974reaching,ren2007information,olfati2007consensus}. Each agent has a local guess of the information state, which has to be updated according to some rule, typically a function of the information states of neighouring agents. Consensus is achieved if all the information states converge to the same value. 
	Achieving consensus is an essential task in many distributed control scenarios where a number of control units (``agents'') interact to achieve a common aim. 
	Consensus problems in  multi-agent systems require 
	the agents  to reach an agreement over a certain real-valued
	scalar or vector, e.g., 
	\cite{degroot1974reaching,ren2007information,olfati2007consensus}. Each
	agent has a local guess of this entity, called the agent's information state, which has to be
	updated according to some rule, typically a function of the
	information states of neighbouring agents. Consensus is achieved if all
	the information states converge to the same value.	
	Consensus-based approaches have been proven to be valuable choices in a wide set of problems, as, for example, the rendez-vous problem \cite{martinez2005robust}, control of vehicle formation \cite{fax2004information}, or the so-called flocking problem \cite{olfati2006flocking}.
	Classical approaches consider \textit{communication} and \textit{computation} as two distinct tasks. Indeed, as communication strategies are usually designed to reliably deliver each information state to a subset of agents by creating independent communication channels, agents have knowledge of other agents' information states.
	%each controller assumes the (perfect) knowledge of the information states that it is supposed to receive.
	%are supposed to be delivered to this controller.
	In general, however, each agent is only interested in a function of other agents' information states, which carries less information (in the information-theoretic sense) than the knowledge of individual information states. This opens the door to significant performance gains.
	%by "shifting" some computational tasks of the controller to the underlying wireless communication, thereby merging communication and computation. 
	Inspired by \cite{stanczak2009fundamentals}, the authors of \cite{goldenbaum2012nomographic} proposed an approach that merges communication and computation of nonlinear functions and is based on the nomographic representation of functions. As pointed out in \cite{goldenbaum2012nomographic}, Buck proved in \cite{buck1979approximate} that every real-valued multivariate function is representable in its nomographic form as a function of a finite sum of univariate functions. Based on this deep insight, the authors of \cite{goldenbaum2012nomographic} concluded that the superposition property of the wireless channel (also called the broadcast property) can be used to approximate an arbitrary function of transmitted signals. 
	According to
	this, each agent simultaneously broadcasts a suitably chosen function
	of its information state. Then, each agent postprocesses the received
	signal, which is a noisy superposition of the locally preprocessed
	information states transmitted by its neighbours, to estimate the
	desired function value.	
	If the goal is to achieve average consensus, the employed consensus
	function is typically linear in the neighbouring agents' information
	states. In this case, the consensus function is already expressed in
	its nomographic representation with both pre- and postprocessing
	functions continuous in the set of real numbers
	\cite{goldenbaum2012nomographic}.  Using the superposition property of wireless channels then allows for significantly faster convergence when compared to standard communication protocols, %  wireless channelHarnessing wireless interference in
	% order to merge communication and computation for an average consensus
	% problem brings about a faster convergence and more efficient solution,
	but introduces  distortions (namely, the unknown channel coefficients)
	proportional to the transmitted signals, which, if not properly addressed,  will cause undesired
	behaviour.
	Existing approaches to exploit superposition neglected the influence of channel coefficients, by considering ideal MACs (wireless multiple access channels) \cite{goldenbaum2012nomographic,zheng2012fast}. We will relax these assumptions and assume, in a realistic way, channel coefficients with no constraints apart from positivity. 
	%A stabilizing robust controller that reaches consensus will be proposed, but the cost to pay for faster convergence is that the resulting consensus value will depend on the channel coefficients.
	
	An outline of this paper is as follows. In \Cref{sec:model}, consensus problems on graphs are presented; channel superposition and usage of interference for consensus problems over wireless networks are then explained. In \Cref{sec:ctrlDesign}, a consensus algorithm exploiting superposition is proposed. The influence of its parameters on convergence rate and consensus value is addressed in \Cref{sec:influence} and illustrated via simulations in \Cref{sec:numEx}. Finally, in \Cref{sec:conclusion}, concluding remarks are stated.
	
	\subsection*{Nomenclature} 
	\theoremstyle{definition}
	We use $\mathbb{N}$ and $\mathbb{R}$ to denote, respectively, the set of positive integers and the set of real numbers. The set of positive real numbers and nonnegative real numbers are denoted, respectively, by $\mathbb{R}_{>0}$ and $\mathbb{R}_{\geq0}$. Given a scalar $a$, its absolute value is denoted by $|a|$. The closed unit interval is $\mathbb{E}:=[0,1]\subset \mathbb{R}$.
	The $n\times m$ zero matrix is denoted by $\mathbf{0}_{n\times m}$.
	Given a matrix $A$, its transpose is $A'$, while its conjugate transpose is $A^*$. The trace of a matrix $A$ is denoted by $tr(A)$. The element in position $ (i,j) $ of $A$ is referred to as $[A]_{ij}$. The $n\times m$ matrix $A$ is positive (nonnegative), denoted by $A>0$ ($A\geq 0$), if $\forall i,j : 1\leq i\leq n,\ 1\leq j\leq m$, $[A]_{ij}>0$ ($[A]_{ij}\geq0$). $A\geq0$ is row-stochastic if $A\mathbf{1}=\mathbf{1}$ where $\mathbf{1}$ is the column vector with all ones. 
	Two $n\times n$ nonnegative matrices $A$ and $B$ are of the same type (denoted by $A\sim B$) if they have zero
	entries in the same locations.
	$A$ is double-stochastic if $A$ and $A'$ are both row-stochastic. A nonnegative square matrix $A$ is said to be primitive if there exists $k\in\mathbb{N}$ such that $A^k>0$.
	Eigenvalues of the $n\times n$ matrix $A$ are denoted by $\lambda_i(A)$, $1\leq i\leq n$, and assumed without loss of generality to be ordered as follows:  $|\lambda_1(A)|\leq|\lambda_2(A)|\leq\dots\leq|\lambda_n(A)|$. The identity matrix of dimension $n\times n$ is denoted by $\mathbb{I}_{n}$; usually, in the event that the context is clear, subscripts are neglected. 
	
	The convex hull $\mathcal{C}(\mathbf{S})$ of a set $\mathbf{S}=\{\mathbf{v}_i\in\mathbb{R}^n,\ 1\leq i\leq m,\ m\in\mathbb{N},\  n\in\mathbb{N}\}$ is the intersection of all convex sets containing $\mathbf{S}$. So we have $\mathcal{C}(\mathbf{S})=\{\sum_{j=1}^m \lambda_j \mathbf{v}_j:\lambda_j\geq0\ \ \forall j, \ \sum_{j=1}^m \lambda_j=1\}$. 
	
%	In the following, we will consider the convex hull of a vector, which is defined as above, but with $n=1$.
		
	%The convex hull of a set $\mathbf{S}=\{\mathbf{v}_i\in\mathbb{R}^n,\ 1\leq i\leq m,\ m\in\mathbb{N},\  n\in\mathbb{N}\}$ is the intersection of all convex sets containing $\mathbf{S}$, and it is denoted by $\mathcal{C}(\mathbf{S})$, i.e. $\mathcal{C}(\mathbf{S})=\{\sum_{j=1}^m \lambda_j \mathbf{v}_j:\lambda_j\geq0\ \ \forall j, \ \sum_{j=1}^m \lambda_j=1\}$.
	
	Given a discrete-time signal $p(k):\mathbb{N}\rightarrow\mathbb{R}$, its zeta-tranform is denoted by $P(z)=\mathcal{Z}(p(k))$. 
	
	Finally, given a finite set $\mathcal{V}$, its cardinality is denoted by $|\mathbf{\mathcal{V}}|$.
	\begin{definition}[Directed graph]
		A directed graph (or digraph) is a pair $(\mathcal{N},\mathcal{A})$, where $\mathcal{N}$ represents a finite set of nodes and $\mathcal{A}\subseteq\mathcal{N}\times\mathcal{N}$ is the set of arcs. 
	\end{definition}
	In the following, we always assume that $(i,i)\not\in\mathcal{A}$, $\forall i\in\mathcal{N}$.
	\begin{definition}[Neighbors]
		Given a directed graph $(\mathcal{N},\mathcal{A})$, the set of neighbours of a node $i\in\mathcal{N}$, denoted by ${N}_i$, is the set of those nodes $l\in\mathcal{N}$ for which $(l,i)\in\mathcal{A}$. 
	\end{definition}
	By the assumption above, $i\not\in{N}_i$. 
	\begin{definition}[Weighted directed graph]
		\label{def:wDG}
		A weighted directed graph is a triple $(\mathcal{N},\mathcal{A},w)$, where $(\mathcal{N},\mathcal{A})$ is a digraph and $w:\mathcal{A}\rightarrow\mathbb{R}_{>0}$ associates each arc $(j,i)\in\mathcal{A}$ with a positive weight $w_{ij}$.
	\end{definition} 
	The digraph is balanced if $\forall i\in\mathcal{N}$, $\sum_{j\in {N}_i}w_{ij}=\sum_{\{j:\ i\in N_{j}\}}w_{ji}$, i.e. if, for each node, the sum of the weights of all incoming arcs equals the sum of the weights of all outgoing arcs.
	A directed path in a digraph is a sequence of nodes in which there is an arc pointing from each node in the sequence to its successor in the sequence. 
	The digraph is called strongly connected if there exists a directed path between any two distinct nodes.
	The digraph is called fully connected (or complete) if there exists an arc between any two distinct nodes.
%	\begin{definition}[Hadamard product]
%		The Hadamard product of two matrices $A$ and $B$ of equal dimensions, also referred to as element-by-element product, is the matrix $C=A\circ B$ with $[C]_{ij}=[A]_{ij}[B]_{ij}$. 
%	\end{definition}
%	Given a vector $\mathbf{v}\in\mathbb{R}^n$, the $n\times n$ diagonal matrix $V=\text{diag}(\mathbf{v})$ can be written as a Hadamard product, i.e. $\text{diag}(\mathbf{v})=\mathbb{I}_{n}\circ\mathbf{1}\mathbf{v}'$.
	\section{System model and problem statement}
	%First, we present the consensus problem over weighted digraphs, which will be of use later on since it properly models the impact of nonidealities due to channel superposition.
	\label{sec:model}
	\subsection{Consensus for weighted digraphs}
	\label{subsec:model_cons}
	We consider a time-varying network described by a sequence of weighted directed graphs 
	\begin{equation}
		\Gamma=\{\Gamma_k:\Gamma_k=(\mathcal{N},\mathcal{A},w(k)),\ k\in\mathbb{N}\}
	\end{equation} 
	with $n=|\mathcal{N}|$ communicating agents (nodes) and with a strongly connected topology. Each agent has the following discrete-time integrator dynamics:
	\begin{equation}
		\label{eq:system}
		x_i(k+1)=x_i(k)+u_i(k), 
		\
		i\in\{1,\dots,n\}
		.
	\end{equation}
	$x_i:\mathbb{N}\rightarrow\mathbb{R}$ is the agent's state and $u_i:\mathbb{N}\rightarrow\mathbb{R}$ its input. The system (\ref{eq:system}) can be also expressed compactly as
	\begin{equation}
		\label{eq:system_matrix}
		\mathbf{x}(k+1)=\mathbf{x}(k)+\mathbf{u}(k),
	\end{equation}
	where $\mathbf{x}(k)=[x_1(k),\dots,x_n(k)]'$ and $\mathbf{u}(k)=[u_1(k),\dots,u_n(k)]'$, $\forall k\in\mathbb{N}$.
%	A commonly used \cite{ren2008distributed} linear consensus protocol is 
%	\begin{equation}
%		\label{eq:controlInput}
%		u_i(k)=\epsilon_k\sum_{j\in {N}_i}w_{ij}(k)(x_j(k)-x_i(k)),
%	\end{equation}
%	where, according to \cite{olfati2007consensus}, $\epsilon_k\in \left(0,\Delta(k)\right)$ with
%	\begin{equation}
%		\label{eq:delta_graph}
%		\Delta(k)=\frac{1}{\text{max}_i(\sum_{j\in {N}_i} w_{ij}(k))}\ \ \ \ \ \forall k\in\mathbb{N}.
%	\end{equation}
	\begin{definition}[Perron matrix]
		\label{def:perron}
		Let a graph $\Gamma_k=(\mathcal{N},\mathcal{A},w(k))$, consisting of $n$ communicating agents with dynamics (\ref{eq:system}), and a parameter $\epsilon_k\in \left(0,\Delta_k\right)$ with
		\begin{equation}
			\label{eq:delta_graph}
			\Delta_k=\frac{1}{\text{max}_i(\sum_{j\in {N}_i} w_{ij}(k))}
		\end{equation}
		be given. The \textit{Perron matrix} of $\Gamma_k$ with parameter $\epsilon_k$ is the matrix $D_n(k)$ defined to be
		\begin{equation}
			\label{eq:perron_def}
			D_n(k):=\mathbb{I}_n-\epsilon_k \mathcal{L}(\Gamma_k),
		\end{equation}
		where $\mathcal{L}(\Gamma_k)$ is the Laplacian of $\Gamma_k$ \cite{ren2007information}. The entries of $D_n(k)$ are $[D_n(k)]_{ii}=1-\epsilon_k\sum_{j\in {N}_i}w_{ij}(k)>0$, $\forall i\in\mathcal{N}$, $[D_n(k)]_{ij}=\epsilon_k w_{ij}(k)$, $\forall (j,i)\in\mathcal{A}$, and $[D_n(k)]_{ij}=0$, $\forall i\not=j$ with $(j,i)\not\in\mathcal{A}$.
		We refer to Lemma 3 in \cite{olfati2007consensus} for the properties of the \textit{Perron matrix}, which is row-stochastic by construction, and primitive if $\Gamma_k$ is strongly connected. 
	\end{definition}
	In literature (see e.g. \cite{ren2008distributed}), the linear consensus protocol 
	\begin{equation}
		\label{eq:controlInput}
		u_i(k)=\epsilon_k\sum_{j\in {N}_i}w_{ij}(k)(x_j(k)-x_i(k))
	\end{equation}
	is widely used and can be expressed in matrix form as
	\begin{equation}
		\label{eq:controlInputMatForm}		
		\mathbf{u}(k)=-\epsilon_k \mathcal{L}(\Gamma_k)\mathbf{x}(k).
	\end{equation}
	
	By applying (\ref{eq:controlInputMatForm}) to the system (\ref{eq:system_matrix}), the closed loop dynamics becomes
	\begin{equation}
		\label{eq:sysUnbal}
		\mathbf{x}(k+1)=D_n(k)\mathbf{x}(k).
	\end{equation}
%	
%	Equation (17) in \cite{olfati2007consensus} shows that
%	\begin{equation}
%		\label{eq:eigenD_L}
%		\lambda_{n+1-i}(D_n(k))=1-\epsilon_k\lambda_i(L(k))\ \ \ i=1,\dots,n	,
%	\end{equation}
%	since the eigenvectors of $D_n(k)$ and $L(k)$ are the same. 
	
%	Let's refer to Lemma 3 in \cite{olfati2004consensus} for the properties of the Perron matrix $D_n(k)$, which is row-stochastic, therefore $\rho(D_n(k))=\max\limits_i|\lambda_i(D_n(k))|=\lambda_1(D_n(k))=1$ as the largest eigenvalue, which is also simple. 
	
	As mentioned in \Cref{def:perron}, $D_n(k)$ is primitive, therefore it has a unique real eigenvalue that strictly dominates the moduli of all other eigenvalues, which is $\rho(D_n(k))=\lambda_n(D_n(k))=1$ since $D_n(k)$ is also row-stochastic. By the\textit{ Perron-Frobenius theorem},
	in case of a time-invariant problem (i.e. $\forall k\in\mathbb{N}$, $w(k)=w$) with graph $\Gamma_k$ unbalanced, the consensus value will be $x^*=\mathbf{w}'\mathbf{x}(0)$, where $\mathbf{w}'D_n=\mathbf{w}'$ and $\mathbf{w}'\mathbf{1}=1$. Accordingly, $x^*\in\mathcal{C}(\mathbf{x}(0))$. In case of $\Gamma_k$ balanced, $D_n$ is double stochastic and consequently $x^*=\frac{1}{n}\mathbf{1}'\mathbf{x}(0)$, which is (linear) average consensus.
	
	In the general case, some convergence results for time-variant multi-agent systems are presented in \cite{moreau2005stability}. 
	In the case considered in this paper, $\Gamma$ is a sequence of weighted digraphs with the same topology but with different positive weights. By \Cref{def:perron}, for a strongly connected topology, $D_n(k)$ will be a sequence of row-stochastic primitive matrices of the same type and with positive diagonal entries. Their product is characterized by the following result.
	\begin{proposition}
		\label{prop:productPrimitive}
		Given two nonnegative $n\times n$ primitive matrices $A,B$ with positive diagonal entries, then $AB$ will be primitive with positive diagonal entries.
		
		\begin{proof} 
			By \cite[p. 3]{seneta2006non}, if $A$ is primitive, any nonnegative matrix $\tilde{A}$ of the same type as $A$ is primitive. If $\tilde{A}$ is primitive and $C$ is nonnegative, $(\tilde{A}+C)$ is primitive.
			As
			\begin{equation}
				[AB]_{ij}=\sum_{k=1}^{n} [A]_{ik}[B]_{kj}\geq [A]_{ij}[B]_{jj},
			\end{equation}
			and $[B]_{jj}>0$, $[AB]_{ij}$ is positive whenever $[A]_{ij}>0$. We can therefore write
			\begin{equation}
				AB = \tilde{A}+C,
			\end{equation}
			where $\tilde{A}$ is nonnegative and of the same type as $A$, and $C$ is nonnegative. Hence, with the above argument, $AB$ is primitive. Positivity of its diagonal elements is straightforward as $[AB]_{ii}=\sum_{k=1}^{n}[A]_{ik}[B]_{ki}\geq [A]_{ii}[B]_{ii}>0$.
			
%			
%			Given two primitive matrices $A$ and $B$, each entry of their product $AB$ is
%			\begin{equation}
%				[AB]_{ij}=\sum_{k=1}^{n} [A]_{ik}[B]_{kj}.
%			\end{equation}
%			If $[A]_{ij}>0$ and $[B]_{jj}>0$ (positive diagonal), then $[AB]_{ij}>0$. Conversely, if $[A]_{ij}=0$, $[AB]_{ij}\geq0$. By this, it exist $C\geq0$, such that $AB$ will be a nonnegative matrix of the same type as $(A+C)$, therefore primitive.
			\qed 
		\end{proof}
	\end{proposition}
	By \Cref{prop:productPrimitive}, $\forall k,h\in\mathbb{N}$, the product $D_n(k+h)D_n(k+h-1)\dots D_n(k)$ results in a primitive and row-stochastic matrix, therefore, according to \cite{wolfowitz1963products}, (\ref{eq:sysUnbal}) achieves consensus.
	%and according to Wolfowitz (), since $\forall k\in\mathbb{N}$, the product $D_n(k+1)D_n(k)$ is primitive and row-stochastic, (\ref{eq:sysUnbal}) achieves consensus.
	%However, since $\Gamma$ is a sequence of weighted digraphs with the same topology (but different positive weights), by \cite{ren2007information} p. 76, (\ref{eq:sysUnbal}) achieves consensus if this topology is strongly connected.	
	%	On the other hand, if the time-variant system (\ref{eq:sysUnbal}) is given, \cite{moreau2005stability} proved that consensus is globally asymptotically reached, provided that the union of communication graphs $\Gamma_k\in\Gamma$ is strongly connected for all $k$.	
	%granted that $D_n(k)\in\mathcal{D}$, where $\mathcal{D}$ is the set of Perron matrices associated with $\Gamma$, which is assumed to be ultimately connected through time \cite{olfati2007consensus}. 
	In addition, the agreement value lies in the convex hull of the initial states.
	%Still $x^*\in\mathcal{C}(\mathbf{x}(0))$. 
	Some examples of average-consensus for time-variant systems can be found in \cite{sun2008average}.
	Next we present a result that is used later in this paper.
	\begin{proposition}
		\label{prop:solveGraphExist}
		Given a row-stochastic matrix $P$ of dimension $n$ and a scalar $\epsilon\in\mathbb{R}_{>0}$, there exists a directed graph $\Gamma_P$, such that $P$ is the Perron matrix of $\Gamma_P$ with parameter $\epsilon$. Moreover, if $P$ is positive, $\Gamma_P$ is fully connected. If $P$ is symmetric, $\Gamma_P$ is undirected.
		
		\begin{proof}
			By (\ref{eq:perron_def}) we need to show that, given $P$ and $\epsilon\in\mathbb{R}_{>0}$, there is an adjacency matrix $A\in\mathbb{R}_{\geq0}^{n\times n}$ of a graph with $tr(A)=0$, such that $P=I-\epsilon(D-A)$, where $D$ is the degree matrix of $A$ as defined in \cite{olfati2004consensus}, with $[D]_{ii}=\sum_{j=1}^{n}[A]_{ij}$, $1\leq i\leq n$. This condition can be written as
			\begin{equation}
			\label{eq:solveGraphExist}
				\begin{cases}
					[P]_{ii}=1-\epsilon\sum_{i\not=j}a_{ij} &\forall i\\
					[P]_{ij}=\epsilon a_{ij} &\forall i\not= j\\
					\sum_j [P]_{ij}=1 &\forall i					
				\end{cases}.
			\end{equation}
			By (\ref{eq:solveGraphExist}), $A$ is an adjacency matrix of a weighted digraph, whose elements are $a_{ij}=\frac{[P]_{ij}}{\epsilon}$, $\forall i\not=j$, $1\leq i\leq n$, $1\leq j\leq n$, and $a_{ii}=0$, $1\leq i\leq n$. 
			Whenever P is positive, then, $a_{ij}>0$, $\forall i\not=j$, in which case $\Gamma_P$ is fully connected. If $P$ is symmetric, we have  $a_{ij}=\frac{[P]_{ij}}{\epsilon}=\frac{[P]_{ji}}{\epsilon}=a_{ji}$, $\forall i\not=j$,  $1\leq i\leq n$, $1\leq j\leq n$, so that $\Gamma$ is undirected, since its adjacency matrix is symmetric.
			This completes the proof.
			\qed
		\end{proof}
	\end{proposition}
	\subsection{Exploiting channel superposition for average consensus}
	To introduce the underlying idea, we first review some existing results. By \cite{buck1979approximate}, we know that each multivariate function $f:\mathbb{E}^n\rightarrow\mathbb{R}$ has a nomographic representation:
	\begin{equation}
		\label{eq:nomog}
		f(x_1,\dots,x_n)=\psi(\sum_{j=1}^n \phi_j(x_j)), 
	\end{equation}
	for some $\psi:\mathbb{R}\rightarrow\mathbb{R}$ and $\phi_j:\mathbb{E}\rightarrow\mathbb{R}$. We are interested in nomographic representations since they allow us to exploit the superposition (interference) property of the wireless channel for function computation over a multi-agent wireless network.
%	Distortions in the transmitted signals and further non-idealities will be described according to the upcoming model.
%	
%	In order to merge \textit{communication} and \textit{computation}, as intended in \cite{goldenbaum2013harnessing}, we make use of the broadcast property of the wireless channel. However, to present this idea, we have to mention some mathematical results. 
%	According to \cite{buck1979approximate}, each multivariate function $f:\mathbb{E}^n\rightarrow\mathbb{R}$ can be expressed as a function of a finite sum of univariate functions, that is to say $f(x_1,\dots,x_n)=\psi(\sum_{j=1}^n \phi_j(x_j))$, with $\psi:\mathbb{R}\rightarrow\mathbb{R}$ and $\phi_j:\mathbb{E}\rightarrow\mathbb{R}$. 
%	%Function $\psi$ depends on $f$, whilst $\phi_j$ can be chosen to be independent on $f$. 
	
	We consider a wireless network represented by a directed graph $(\mathcal{N},\mathcal{A})$, where $\mathcal{N}=\{1,\dots,n\}$ is the set of nodes and $(i,j)\in\mathcal{A}\subset\mathcal{N}\times\mathcal{N}$ if and only if information transmission from $i\in\mathcal{N}$ to $j\in\mathcal{N}$ is established. 
	
	If, after the implementation of a general consensus protocol, each agent evolves according to
	\begin{equation}
		x_i(k+1)=f_i(x_1(k),\dots,x_n(k)), \ i\in\mathcal{N},
	\end{equation}
	%As in the spirit of \cite{goldenbaum2013harnessing}, 
	it does not need to reconstruct the individual information states of other agents.
	By using a nomographic representation of $f_i$ and the interference property of the wireless channel, one use of the noiseless channel is sufficient to compute the function.
	For this, each node $i\in\mathcal{N}$ will broadcast simultaneously with all the other nodes at instant $k\in\mathbb{N}$ its pre-processed information state $\phi_i(x_i(k))$. 
	By describing the communication with the standard affine model of a wireless multiple-access channel (MAC) \cite{goldenbaum2012nomographic}, the real-valued signal received at node $i\in\mathcal{N}$ becomes
	\begin{equation}
		\label{eq:receivedNom}
		Y_i(k)=\sum_{j\in {N}_i} h_{ij}(k)\phi_j(x_j(k))+v_i(k),
	\end{equation}	
	where $h_{ij}(k)\in\mathbb{R}_{>0}$, $\forall j\in N_i$ (otherwise $0$), denotes a channel coefficient from node $j$ (transmitter) to node $i$ (receiver) and $v_i\in\mathbb{R}$ is the corresponding receiver noise.
	Ideally, $h_{ij}(k)=1$, $\forall j \in N_i$, and $v_i(k)=0$, $\forall i \in\mathcal{N}$. In this ideal case, every node $i\in\mathcal{N}$ computes $x_i(k+1)=\psi_i(\phi_i(x_i(k)) + Y_i(k))=f_i(x_1(k),\dots,x_n(k))$.
	
	To achieve average-consensus, each node $i\in\mathcal{N}$ may compute at iteration $k$
	\begin{equation}
		x_i(k+1)=f_i(x_1(k),\dots,x_n(k))=\frac{\sum_{j\in N_i\cup\{i\}} x_j(k)}{|N_i|+1}.
	\end{equation}
	This function is nomographic with $\phi_j(y_j)=y_j$ and $\psi_i(y)=\frac{y}{|N_i|+1}$, both trivially continuous and differentiable in $\mathbb{R}$. 
	If the receiver noise can be neglected (i.e. $\forall k\in\mathbb{N}$, $\forall i\in\mathcal{N}$, $v_i(k)=0$), then, for the average-consensus problem,  (\ref{eq:receivedNom}) becomes
	%For the average-consensus problem, (\ref{eq:receivedNom}) becomes
	\begin{equation}
		\label{eq:Yik}
		Y_i(k)=\sum_{j\in {N}_i} h_{ij}(k)x_j(k).
	\end{equation}	
	%where the receiver noise has been neglected (i.e. $\forall k\in\mathbb{N}$, $\forall i\in\mathcal{N}$, $v_i(k)=0$).
	Each agent $i\in\mathcal{N}$ then computes $x_i(k+1)=\frac{1}{|N_i|+1}(x_i(k)+Y_i(k))$.
	Therefore 
	\begin{equation}
		\mathbf{x}(k+1)=D(k)\mathbf{x}(k)
	\end{equation}
	with
%	With this communication model in hand, the dynamics matrix of the system $\mathbf{x}(k+1)=D(k)\mathbf{x}(k)$ becomes
	\begin{equation}
		\label{eq:matrixNomo}
		D(k)=
		\begin{bmatrix}
			\frac{1}{|N_1|+1}	&\frac{h_{12}(k)}{|N_1|+1}	&\dots	&\frac{h_{1n}(k)}{|N_1|+1}\\
			\frac{h_{21}(k)}{|N_2|+1}	&\frac{1}{|N_2|+1}	&\dots	&\frac{h_{2n}(k)}{|N_2|+1}\\
			\dots &\dots &\dots &\dots\\
			\frac{h_{n1}(k)}{|N_n|+1}	&\frac{h_{n2}(k)}{|N_n|+1}	&\dots	&\frac{1}{|N_n|+1}			
		\end{bmatrix}.
	\end{equation}
	$D(k)$ is a sequence of nonnegative square matrices (according to the definition of channel coefficients). 
%	Any off-diagonal entry $[D(k)]_{ij}$, $i\not= j$, is zero if and only if $(j,i)\not\in\mathcal{A}$. 
%	By this, the presence of positive channel coefficients does not change the communication topology. 
	In what follows, the following assumption is made:
	\begin{assumption}
		\label{ass:ass1}
		The communication topology $(\mathcal{N},\mathcal{A})$ is strongly connected.
	\end{assumption}
%	Then, for any set of positive channel coefficients, by \cite{olfati2007consensus}, since the underlying communication graph is strongly connected, $D(k)$ is a primitive matrix. 
	However, $D(k)$ is in general not row-stochastic. Hence, we cannot expect consensus.
	
%	By \cite{olfati2007consensus}, in order to reach a consensus, we need only to verify that $D(k)\in\mathcal{D}$, where $\mathcal{D}$ is used to denote the set of row-stochastic matrices. By \cite{moreau2005stability}, this, together with \Cref{ass:ass1}, guarantees that the time-variant system achieves consensus. 
%	
%	On the other hand, for the time-invariant case,
%	\begin{equation}
%	\mathbf{x}(k+1)=D\mathbf{x}(k), 
%	\end{equation}
%	where $D=D(k)$ for some arbitrary $k$, a primitive row-stochastic $D$ guarantees convergence to consensus.
%	
%	Accordingly, both for time-variant and -invariant case, we search for a controller that makes the dynamics matrix positive row-stochastic.
	
	\section{Controller Design}
	\label{sec:ctrlDesign}
	%
	%
	%
	%
	%
	
	%	In general, $D(k)$ in (\ref{eq:matrixNomo}) is not row-stochastic.
	%	In the following, we present one possible strategy that makes the dynamics row-stochastic. 
	In the following, we discuss a control strategy that achieves consensus despite the a priori unknown channel coefficients.
	\subsection{Protocol design}
	Under the assumption of a noiseless channel, each agent $i\in\mathcal{N}$ broadcasts two orthogonal signals, $\tau_i(k)=x_i(k)$ and $\tau_i'(k)=1$ (by using a MAC of order 2 \cite{goldenbaum2013harnessing}). Due to the superposition property, each agent $i\in\mathcal{N}$ receives from neighbouring agents two orthogonal real-valued signals, $Y_i(k)=\sum_{j\in {N}_i}h_{ij}\tau_{j}(k)$, which is equal to (\ref{eq:Yik}), and
	\begin{equation}
	Y_i'(k)=\sum_{j\in {N}_i} h_{ij}(k)\tau_j'(k)=\sum_{j\in {N}_i} h_{ij}(k).
	\end{equation}
	%under the assumption that channel coefficients $h_{ij}(k)$ remain constant during every step $k\in\mathbb{N}$ for transmissions to agent $i\in\mathcal{N}$ from agent $j\in {N}_i$.
	\subsection{Controller design}
	A controller can then be defined by 
	\begin{equation}
		u_i(k)=\sigma_i\left(\frac{Y_i(k)}{Y_i'(k)}-x_i(k) \right)
	\end{equation}
	resulting in
	\begin{equation}
		\label{eq:eq_stabil}
		x_i(k+1)=(1-\sigma_i)x_i(k)+\sigma_i\left[\frac{Y_i(k)}{Y_i'(k)}\right],
	\end{equation}
	where $\sigma_i\in(0,1)$, $1\leq i\leq n$. As the quantity $\frac{Y_i(k)}{Y_i'(k)}$ is a weighted average of neighbours' information states,
	\begin{equation}
		\frac{Y_i(k)}{Y_i'(k)}=\frac{\sum_{j\in {N}_i}h_{ij}(k)x_{j}(k)}{\sum_{j\in {N}_i}h_{ij}(k)},
	\end{equation}
	the system can be written in matrix form as
	\begin{equation}
		\label{eq:systemSigma}
		\mathbf{x}(k+1)=D_n^\sigma(k) \mathbf{x}(k),
	\end{equation}
	where
	
	{\small	
	\begin{equation}
		\label{eq:matrixSigma}
		D_n^\sigma(k)=
		\begin{bmatrix}
			(1-\sigma_1)	&\frac{\sigma_1 h_{12}(k)}{\sum_{j\in N_1}h_{1j}(k)}	&\dots	&\frac{\sigma_1 h_{1n}(k)}{\sum_{j\in N_1}h_{1j}(k)}\\
			\frac{\sigma_2 h_{21}(k)}{\sum_{j\in N_2}h_{2j}(k)}	&(1-\sigma_2)	&\dots	&\frac{\sigma_2 h_{2n}(k)}{\sum_{j\in N_2}h_{2j}(k)}\\
			\dots &\dots &\dots &\dots\\
			\frac{\sigma_n h_{n1}(k)}{\sum_{j\in N_n}h_{nj}(k)}	&\frac{\sigma_n h_{n2}(k)}{\sum_{j\in N_n}h_{nj}(k)}	&\dots	&(1-\sigma_n)
		\end{bmatrix}.
	\end{equation}
	}
	The sequence $D_n^\sigma(k)$ is composed of row-stochastic matrices $\forall k\in\mathbb{N}$. 
	We show next that, $\forall\sigma_i\in(0,1)$, $1\leq i\leq n$, $D_n^\sigma(k)$ is a sequence of primitive matrices of the same type.
	%{By \Cref{ass:ass1} and Lemma 3 in \cite{olfati2007consensus}, for any set of positive channel coefficients, every $D_n^\sigma(k)$ in the sequence will be primitive.
	%Therefore the system (\ref{eq:systemSigma}) reaches the consensus, i.e. $\mathbf{x}(k)\rightarrow\mathbf{x}^*=\mathbf{1}x^*$ where $x^*\in\mathcal{C}(\mathbf{x}(0))$. 
%	Reaching consensus with the proposed broadcast strategy has proved to be faster than with standard protocols like TDMA \cite{goldenbaum2013harnessing}. 	
	%However, the dynamics matrix $D_n(k)$ is non-symmetric in general; therefore (\ref{eq:systemSigma}) converges to a weighted average consensus, i.e., in general $x^*\not=\frac{1}{n}\sum_{i=1}^{n}x_i(0)$.}	
%	instead of $x^*=\frac{1}{n}\sum_{i=1}^n x_i(0)$, the information state at each node converges to $x^*=\sum_{i=1}^n \lambda_i x_i(0)$ for some $\lambda_i>0$, $1\leq i\leq n$, with $\sum_{i=1}^n \lambda_i=1$. 

	We need first to consider a special case; if $\sigma_i$ in (\ref{eq:eq_stabil}) is 
	\begin{align}
		\label{eq:chooseSigma}
		\sigma_i=\bar{\sigma}_i:=\epsilon_k{\sum_{j\in {N}_i}h_{ij}(k)}\ \ \ \ \ &\forall i\in\mathcal{N}, 
	\end{align}
	where $\epsilon_k$ is chosen as
	\begin{equation}
		0<\epsilon_k<\frac{1}{\text{max}_i(\sum_{j\in {N}_i}h_{ij}(k))},
	\end{equation}
	then
	$D_n^\sigma(k)$ is the Perron matrix with parameter $\epsilon_k$ of a directed weighted graph $\Gamma_{\sigma,k}$ (\Cref{prop:solveGraphExist}), and is denoted by $\bar{D}_n^\sigma(k)$. To be precise, we have
	\begin{equation}
		\Gamma_{\sigma,k}=(\mathcal{N},\mathcal{A},w(k)),\ k\in\mathbb{N},
	\end{equation}
	where the weights are determined by the channel coefficients $w_{ij}(k)=h_{ij}(k)$, $\forall(j,i)\in\mathcal{A}$.
	%[As $D_n^\sigma(k)$ is in general neither symmetric nor double stochastic, $\Gamma_{\sigma,k}$ will be directed and unbalanced.]
	By \Cref{ass:ass1}, by \Cref{def:perron}, and since $h_{ij}(k)>0$ if $(j,i)\in\mathcal{A}$, $\bar{D}_n^\sigma(k)$ is a sequence of primitive row-stochastic matrices. 
	
	In the general case, $\forall\sigma_i\in(0,1)$, $1\leq i\leq n$, the resulting $D_n^\sigma(k)$ is a sequence of nonnegative matrices of the same type as $\bar{D}_n^\sigma(k)$, therefore primitive.
	
	As presented in \Cref{sec:model}, since ${D}_n^\sigma(k)$ is a sequence of row-stochastic primitive matrices of the same type, (\ref{eq:systemSigma}) achieves consensus, i.e. $\mathbf{x}(k)\rightarrow\mathbf{x}^*=\mathbf{1}x^*$ where $x^*\in\mathcal{C}(\mathbf{x}(0))$. 	
	However, $D_n(k)$ is non-symmetric in general; therefore (\ref{eq:systemSigma}) converges to a weighted average consensus, i.e., in general $x^*\not=\frac{1}{n}\sum_{i=1}^{n}x_i(0)$.
	
%	Due to \Cref{prop:solveGraphExist}, since in general $h_{ij}(k)\not=h_{ji}(k)$, $\forall(i,j)\in\mathcal{A}$, $D_n^\sigma(k)$ is not symmetric, thus the graph is directed. Moreover, there is no evidence $D_n^\sigma(k)$ is either double-stochastic, therefore $\Gamma_{\sigma,k}$ is unbalanced \cite{lin2014constrained}.
	
	\section{Influence of $\sigma$, $n$, and $h_{ij}$}
	\label{sec:influence}
	% Channel coefficients are assumed independent and identically distributed.
	In the following, we assume $\sigma_i=\sigma$ in (\ref{eq:eq_stabil}). Thus, the closed loop dynamics (\ref{eq:systemSigma}) is characterized by the matrix
	
%	the dynamics of each agent $i\in\mathcal{N}$ becomes
%	\begin{equation}
%		\label{eq:decCtrl}
%		x_i(k+1)=(1-\sigma)x_i(k)+\sigma \frac{\sum\limits_{j\in {N}_i}h_{ij}(k)x_j(k)}{\sum\limits_{j\in {N}_i}h_{ij}(k)},
%	\end{equation}
%	while the dynamics matrix of the system (\ref{eq:systemSigma}) is then

	{\small	
	\begin{equation}
		\label{eq:matrixSigmaConst}
		D_n^\sigma(k)=
		\begin{bmatrix}
		(1-\sigma)	&\frac{\sigma h_{12}(k)}{\sum\limits_{j\in N_1}h_{1j}(k)}	&\dots	&\frac{\sigma h_{1n}(k)}{\sum\limits_{j\in N_1}h_{1j}(k)}\\
		\frac{\sigma h_{21}(k)}{\sum\limits_{j\in N_2}h_{2j}(k)}	&(1-\sigma)	&\dots	&\frac{\sigma h_{2n}(k)}{\sum\limits_{j\in N_2}h_{2j}(k)}\\
		\dots &\dots &\dots &\dots\\
		\frac{\sigma h_{n1}(k)}{\sum\limits_{j\in N_n}h_{nj}(k)}	&\frac{\sigma h_{n2}(k)}{\sum\limits_{j\in N_n}h_{nj}(k)}	&\dots	&(1-\sigma)
	\end{bmatrix}.
	\end{equation}
	}
	The parameter $\sigma\in(0,1)$ represents a stubbornness index: for small values of $\sigma$, each agent relies more on its current information state than on those of its neighbours, intuitively leading to a slower convergence. 
%	In contrast, larger values of $\sigma$ result in a faster convergence which may exhibit an oscillating behaviour.
	
	%The bigger is $|\mathcal{N}|=n$, the more $\mathbf{x}^*\rightarrow\sum_{i\in\mathcal{N}}\frac{x_i(0)}{n}$. 
%	The larger the network size, the closer the limit of (\ref{eq:systemSigma}) is to the linear average consensus. 
%	In fact, when $n$ (the number of agents in the network) tends to infinity, $D_n^\sigma(k)$ converges to a symmetric matrix, i.e. 
%	\begin{multline}
%		\forall k\in\mathbb{N},\ \forall (i,j)\in\mathcal{A},\ \forall\epsilon>0,\ \exists n_0 \text{  such that:}\\
%		\forall n\geq n_0: \left|[D_n^\sigma(k)]_{ij}-\frac{\sigma}{n-1}\right|\leq\epsilon.
%	\end{multline}
%	Consequently, if $n\rightarrow\infty$, then $x^*$ tends to $\frac{1}{n}\sum_{i=1}^{n}x_i(0)$.
	
%	\begin{equation}
%	\lim\limits_{n\rightarrow\infty}[D_n^\sigma(k)]_{ij}=\lim\limits_{n\rightarrow\infty}\frac{\sigma h_{ij}(k)}{\sum_{l\in {N}_i}h_{il}(k)}\sim\frac{\sigma}{n-1},
%	\end{equation}
%	$\forall (i,j) \in\mathcal{A}$. With a larger network, the system gets closer to linear average consensus. 
	
	In the following, we analyse the time-variant and the time-invariant cases separately.
	\subsection{Time-invariant system}
%	\begin{figure}[t]
%		\centering
%		\includegraphics[width=\columnwidth]{img/polar.eps}
%		\caption{Polar plot of $\Sigma(z)$ for different $\sigma$. The low-pass filtering action of $\Sigma(z)$ is then clear. The smaller $\sigma$, the bigger is the attenuation. Conversely, constant signals are passed as-is, i.e. $\Sigma(1)=1\angle0$, $\forall \sigma$.}
%		\label{fig:polar}
%	\end{figure}
	\label{subsec:timeinvSys}
%	Let the system be
%	\begin{equation}
%	\label{eq:timeInvSys}
%	\forall k \in \mathbb{N}, \ \mathbf{x}(k+1)=D_n^\sigma \mathbf{x}(k)=(D_n^\sigma)^k\mathbf{x}(0),
%	\end{equation}
%	where $D_n^\sigma$ is given by (\ref{eq:matrixSigmaConst}) for some arbitrary $k$. The following result shows how $\sigma$ and the channel coefficients affect the limiting point $\mathbf{x}^*=\mathbf{1}x^*$, where $\mathbf{x}(k+1)=(D_n^\sigma)^k\mathbf{x}(0)\xrightarrow[k\rightarrow\infty]{}\mathbf{x}^*$.
If the channel coefficients do not depend on time,
\begin{equation}
	D_n^\sigma(k)=D_n^\sigma, \ \forall k\in\mathbb{N}.
\end{equation}
Then
\begin{equation}
	\mathbf{x}(k)=\left( D_n^\sigma \right)^k \mathbf{x}(0).
\end{equation}
The following proposition shows how the limit $\mathbf{x}^*=\lim\limits_{k\rightarrow\infty}\mathbf{x}(k)$ depends on the parameter $\sigma$ and the channel coefficients.

	\begin{proposition}
		\label{prop:timeInvDep}
		%Consider the iteration in (\ref{eq:timeInvSys}) and 
		Let $\mathbf{w}$ be the left-eigenvector of $D_n^\sigma$ corresponding to the eigenvalue $\lambda_n=1$. Then, the limit point $\mathbf{x}^*=\mathbf{1}x^*$, with ${x}^*=\mathbf{w}'\mathbf{x}(0)$, is independent of $\sigma$, while depending on the channel coefficients as follows
		\begin{equation}
		\label{eq:sigmainx}
		\begin{cases}
		{x}^*=\sum_{i=1}^n \mathbf{w}_{i}x_i(0)\\
		\mathbf{w}_{i}=\sum_{j\in {N}_i}\frac{\mathbf{w}_{j}h_{ji}}{\sum_{l\in {N}_j}h_{jl}}
		%\\\sum_{i=1}^n \mathbf{w}_{i}=1
		\end{cases}.
		\end{equation}
		
		\begin{proof}
			Let $n\in\mathbb{N}$ and $\sigma\in(0,1)$ be arbitrary but fixed. Since $D_n^\sigma$ is primitive, 
			the Perron-Frobenius theorem states that (Theorem 1.2 in \cite{seneta2006non} p. 9), as $k\rightarrow\infty$,
			\begin{equation}
				\label{eq:PFT}
				(D_n^{\sigma})^k=\lambda_n^k\mathbf{v}\mathbf{w}'+0(k^{m-1}|\lambda_{n-1}^k|),
			\end{equation}
			where $\lambda_n$ and $\lambda_{n-1}$ are, respectively, the largest and the second largest eigenvalues of $D_n^\sigma$ and $m$ is the multiplicity of $\lambda_{n-1}$. Additionally, $\mathbf{v}'\mathbf{w}=1$, where $\mathbf{v}>0$ and $\mathbf{w}>0$ are, respectively, the right and left eigenvectors of $D_n^{\sigma}$, associated with its largest eigenvalue $\lambda_n$. 
			Since $D_n^{\sigma}$ is row-stochastic,  we have $\lambda_n=1$, $|\lambda_{n-1}|<1$, and $\mathbf{v}=\mathbf{1}$. 
			Hence, we have $\mathbf{x}^*=\lim\limits_{k\rightarrow\infty}[(D_n^\sigma)^k\mathbf{x}(0)]=\mathbf{v}\mathbf{w}'\mathbf{x}(0)=\mathbf{1}\mathbf{w}'\mathbf{x}(0)$, from which we conclude the first part of (\ref{eq:sigmainx}).
			
			By the definition of the left eigenvector, we have $\mathbf{w}'D_n^\sigma = \mathbf{w}'$. So $\sum_{j=1}^n \mathbf{w}_{j}[D_n^\sigma]_{ji} = \mathbf{w}_{i} $, $\forall i\in\{1,\dots,n\}$. This, by using the entries of $D_n^\sigma$, becomes 
			\begin{equation}
			\mathbf{w}_i(1-\sigma)+\sum_{j\in {N}_i}\mathbf{w}_{j}\sigma\frac{h_{ji}}{\sum_{l\in {N}_j}h_{jl}}=\mathbf{w}_i,
			\end{equation}
			from which the second part of (\ref{eq:sigmainx}) immediately follows. 
			\qed
		\end{proof}
	\end{proposition}
	
%	The larger the network size, the closer the limit of (\ref{eq:systemSigma}) is to the linear average consensus. 
%	In fact, when $n$ (the number of agents in the network) tends to infinity, $D_n^\sigma$ converges to a symmetric matrix, i.e. 
%	\begin{multline}
%		\forall (j,i)\in\mathcal{A},\ \forall\epsilon>0,\ \exists n_0 \text{  such that:}\\
%		\forall n\geq n_0: \left|[D_n^\sigma]_{ij}-\frac{\sigma}{n-1}\right|\leq\epsilon.
%	\end{multline}
%	Consequently, if $n\rightarrow\infty$, then $x^*$ tends to $\frac{1}{n}\sum_{i=1}^{n}x_i(0)$.

	\subsection{Time-variant system}
	Note that in the time-variant case, the equation for agent $i$ can be rewritten as
	\begin{multline}
		\label{eq:decCtrl_ideal_deviation}
		x_i(k+1)=(1-\sigma)x_i(k)+\frac{\sigma}{|N_i|}\sum_{j\in N_i}x_j(k)+\\
			\frac{\sigma}{|N_i|}\frac{\sum_{j\in N_i}\sum_{l\in N_i}(h_{ij}(k)-h_{il}(k))x_j(k)}{\sum_{j\in N_i}h_{ij}(k)},
	\end{multline}
	where the last term on the right hand side of (\ref{eq:decCtrl_ideal_deviation}) represents the impact of the channel at time $k$. Let $\nu_{ij}(k)$ be
	\begin{equation}
		\label{eq:nu}
		\nu_{ij}(k)=\frac{\sum_{l\in N_i}h_{ij}(k)-h_{il}(k)}{\sum_{l\in N_i}h_{il}(k)}x_j(k),
	\end{equation}
	$\forall (j,i)\in\mathcal{A}$, $\forall k\in\mathbb{N}$, and $\nu_{ij}(k)=0$, $\forall (j,i)\not\in \mathcal{A}$, $\forall k\in\mathbb{N}$. 
	If the channel coefficients are realizations of a stochastic process, independent and identically distributed, we can prove that the expected value of $\nu_{ij}(k)$ is $0$, $\forall k\in\mathbb{N}$, $\forall (j,i)\in\mathcal{A}$.
	
	Putting (\ref{eq:nu}) into (\ref{eq:decCtrl_ideal_deviation}) yields
	\begin{equation}
		x_i(k+1)=(1-\sigma)x_i(k)+\frac{\sigma}{|N_i|}\sum_{j\in N_i}x_j(k)+
		\frac{\sigma}{|N_i|}\sum_{j\in N_i}\nu_{ij}(k),
	\end{equation}
	$\forall i\in\mathcal{N}$, which can be written in matrix form as
	\begin{equation}
		\label{eq:decCtrl_matrix_ideal_dev}
		\mathbf{x}(k+1)=D_{A}^\sigma\mathbf{x}(k)+D_{B}^\sigma\mathbf{\nu}(k),
	\end{equation}
	such that the state vector $\mathbf{x}(k)\in\mathbb{R}^n$ and the state disturbance vector $\mathbf{\nu}(k)\in\mathbb{R}^{n^2}$, with $\nu(k)=[\nu_{11}(k),\dots,\nu_{1n}(k),\nu_{21}(k),\dots,\nu_{2n}(k),\dots,\nu_{nn}(k)]'$.
	The dynamics matrix $D_{A}^\sigma$ is a row-stochastic matrix whose elements are $[D_{A}^\sigma]_{ii}=(1-\sigma)$, $\forall i\in\mathcal{N}$,   $[D_{A}^\sigma]_{ij}=\frac{\sigma}{|N_i|}$, $\forall (j,i)\in\mathcal{A}$, and $0$ elsewhere. 
	The matrix $D_{B}^\sigma\in\mathbb{R}^{n\times n^2}$ is a block-diagonal matrix that can be written as
	\begin{equation}
		\label{eq:DB}
		D_B^\sigma = 
		\begin{bmatrix}
			D_{B,1}^{\sigma} &\mathbf{0}_{1\times n} &\dots &\mathbf{0}_{1\times n} &\mathbf{0}_{1\times n} 	\\
			\mathbf{0}_{1\times n} &D_{B,2}^{\sigma} &\dots&\mathbf{0}_{1\times n} &\mathbf{0}_{1\times n} \\
			\dots &\dots &\dots &\dots &\dots \\
			\mathbf{0}_{1\times n} &\mathbf{0}_{1\times n} &\dots &D_{B,n-1}^\sigma&\mathbf{0}_{1\times n}\\
			\mathbf{0}_{1\times n} &\mathbf{0}_{1\times n} &\dots &\mathbf{0}_{1\times n} &D_{B,n}^\sigma\\
		\end{bmatrix},
	\end{equation}
	where $\forall i\in\mathcal{N}$, $D_{B,i}^\sigma$ is a row-vector of dimension $n$, whose elements are $[D_{B,i}^\sigma]_j=\frac{\sigma}{|N_i|}$, $\forall j \in\{j\mid (j,i)\in\mathcal{A}$ \}, and $0$ elsewhere.
%	whose elements are $[D_{n,B}^\sigma]_{ii}=0$, $\forall i\in\mathcal{N}$,  $[D_{n,B}^\sigma]_{ij}=\frac{\sigma}{|N_i|}$, $\forall (i,j)\in\mathcal{A}$, and $0$ elsewhere. Additionally, $\nu(k)$ can be regarded as the state disturbance at instant $k\in\mathcal{N}$, and $[\mathbf{\nu}(k)]_{ij}=\nu_{ij}(k)$, $\forall k\in\mathbb{N}$, $\forall (i,j)\in\mathcal{A}$, and $0$ elsewhere. The following takes inspiration from the frequency-domain analysis of consensus problems presented in \cite{spanos2005dynamic}. 
	
	We define $X(z)$ and $\mathcal{V}(z)$ as the Zeta-transforms of their respective time-domain signals, i.e. $X(z)=\mathcal{Z}(\mathbf{x}(k))$ and $\mathcal{V}(z)=\mathcal{Z}(\nu(k))$.
	In a complex frequency domain representation, (\ref{eq:decCtrl_matrix_ideal_dev}) becomes
	\begin{equation}
		X(z)=F_A(z)\mathbf{x}(0)+F_B(z)\mathcal{V}(z),
	\end{equation}
	where $F_A(z)=z(z\mathbb{I}_n-D_{A}^\sigma)^{-1}$ and $F_B(z)=(z\mathbb{I}_n-D_{A}^\sigma)^{-1}D_{B}^\sigma$, respectively a n-dimensional square matrix and a $(n\times n^2)$ matrix in the complex frequency domain. 
	In the following, we assume that $\Gamma$ is a sequence of fully connected $\Gamma_k$, then, $\forall k\in\mathbb{N}$, $\forall i\in\mathcal{N}$, $|N_i|=n-1$. 
	%If, for ease of calculation, we keep a fully connected communication graph, $|N_i|=n-1$, $\forall i\in\mathcal{N}$. The poor transmission rate between two agents is modeled with the corresponding channel coefficient. 
	%Consequently, $D_{n,A}^\sigma$ becomes a symmetric row-stochastic matrix, and $D_{n,B}^\sigma$ becomes a symmetric null-diagonal matrix with all off-diagonal elements equal to $\frac{\sigma}{n-1}$. 
	
	We solve $F_A(z)$ and $F_B(z)$ as functions of the complex variable $z$ and of parameters $n$ and $\sigma$; then, the final value theorem for time-discrete systems gives
	\begin{align}
		\lim\limits_{k\rightarrow\infty}\mathbf{x}(k)&=\lim\limits_{z\rightarrow 1}(z-1)\left(F_A(z)\mathbf{x}(0)+F_B(z)\mathcal{V}(z)\right)
		\\		
		\label{eq:limitFreqSys}
		&=\lim\limits_{z\rightarrow 1}
		\frac{1}{n}\mathbf{1}\mathbf{1}'\mathbf{x}(0)+\frac{\sigma}{n(n-1)}\Xi \mathcal{V}(z)
		,
	\end{align}
	$n\in\mathbb{N}$, $\sigma\in(0,1)$ and $\Xi$ a $n\times n^2$ matrix in the form
	\begin{equation}
		\Xi=\mathbf{1}_n \xi',%\underbrace{\left[ \xi_n,\xi_n,...,\xi_n\right]}_{\text{n times}},
	\end{equation}
	where $\xi\in\mathbb{R}^{n^2}$ is a column vector whose elements are $[\xi]_{hn+1}=0$, $\forall h=0\dots n-1$, and $1$ elsewhere.
	
	By (\ref{eq:limitFreqSys}), stubborn systems (small values of $\sigma$) reduce the impact of time-varying channel coefficients on the agreement value more than systems with higher values of $\sigma$. 
	%However, as presented in \Cref{sec:influence}, the convergence rate decreases if $\sigma>0$ becomes smaller.
	However, as argued in the beginning of this section, we may expect the convergence rate to decrease if $\sigma>0$ becomes smaller.
	
	A smaller impact of time-varying channel coefficients is also a benefit of a larger network (where $n$ is large).
	
	The formal analysis of the transient behaviour (function of time-varying channel coefficients), together with the relaxation of the assumption of a fully connected topology, will be the subject of future work. 
	
%	When the channel coefficients are time-varying, $\mathbf{x}^*$ depends on $\sigma$, which can be tuned in order to get closer to linear average consensus. Equation (\ref{eq:decCtrl}) represents a discrete-time low-pass filter \cite{bolzern1998fondamenti}. Constant channel coefficients are passed as is (see \Cref{subsec:timeinvSys}), whilst their variation though time will be attenuated due to the filtering action of $\Sigma(z)$. We can roughly show it by writing (\ref{eq:decCtrl}) in frequency domain as 
%	\begin{equation}
%		X_i(z)=\Sigma(z)\left(\sum\limits_{j\in {N}_i}H_{ij}(z) X_j(z)\right),
%	\end{equation}
%	where $X_i(z)=\mathcal{Z}(x_i(k))$, $\forall i\in\mathcal{N}$, $H_{ij}(z)=\mathcal{Z}\left(\frac{h_{ij}(k)}{\sum_{l\in {N}_i}h_{il}(k)}\right)$, $\forall (i,j)\in\mathcal{A}$, and $\Sigma(z)=\frac{\sigma}{z-(1-\sigma)}$, whose polar diagram is reported in \Cref{fig:polar}. Time-varying channel coefficients introduce in the spectrum of $H_{ij}(z)$ high frequency components, which will be attenuated according to the polar diagram of $\Sigma(z)$. Smaller $\sigma$ enhances the filtering attenuation action.
	
	\section{Numerical example}
	\label{sec:numEx}	
	\begin{figure}[htbp]
		\centering
		\includegraphics[width=.6\columnwidth]{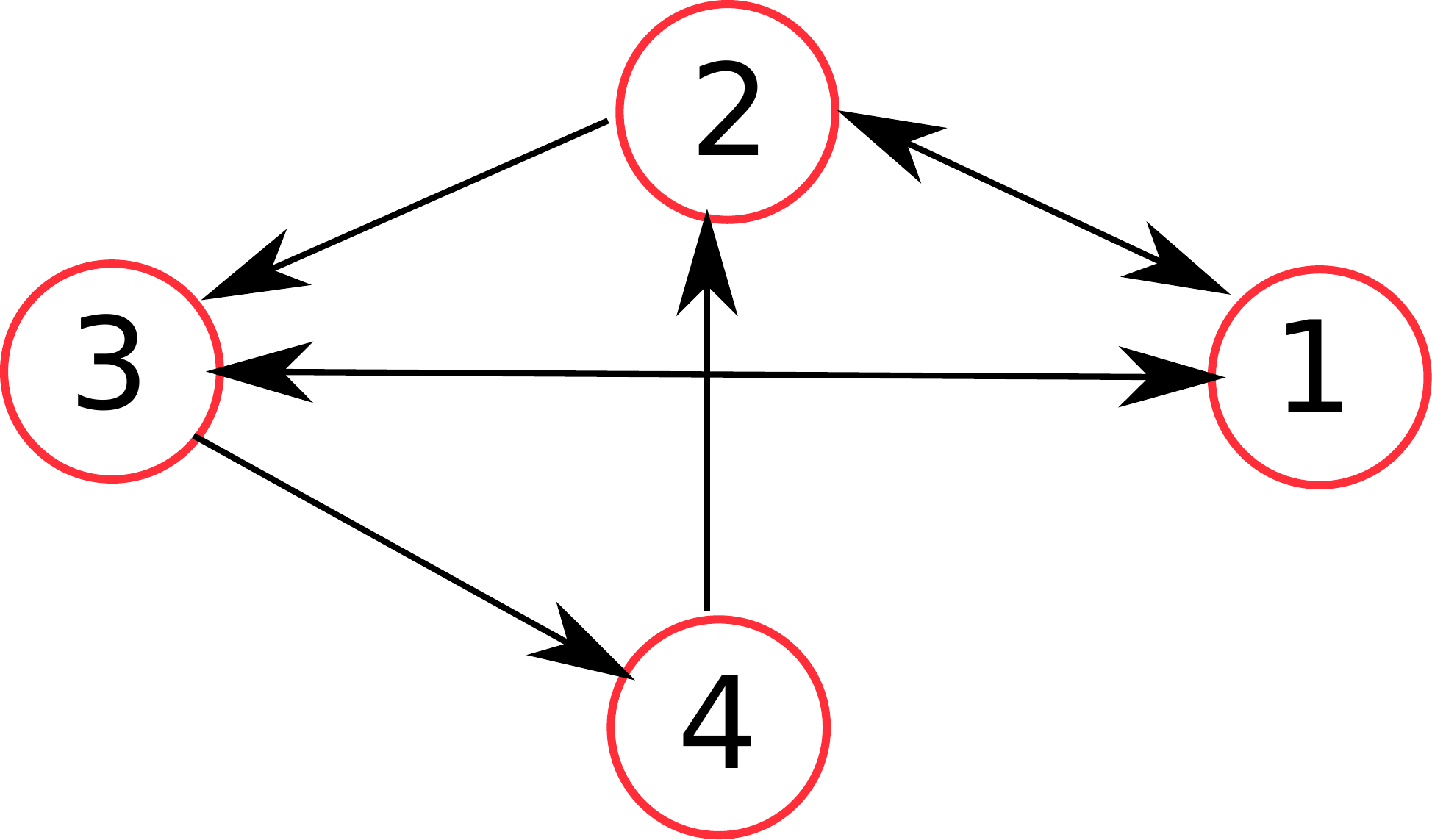}
		\caption{Communication digraph with topology $(\mathcal{N},\mathcal{A})$.}
		\label{fig:comm_graph}
	\end{figure}
\begin{figure}[htbp]
		\centering
		\begin{subfigure}[b]{\columnwidth}
c			\includegraphics[width=\columnwidth]{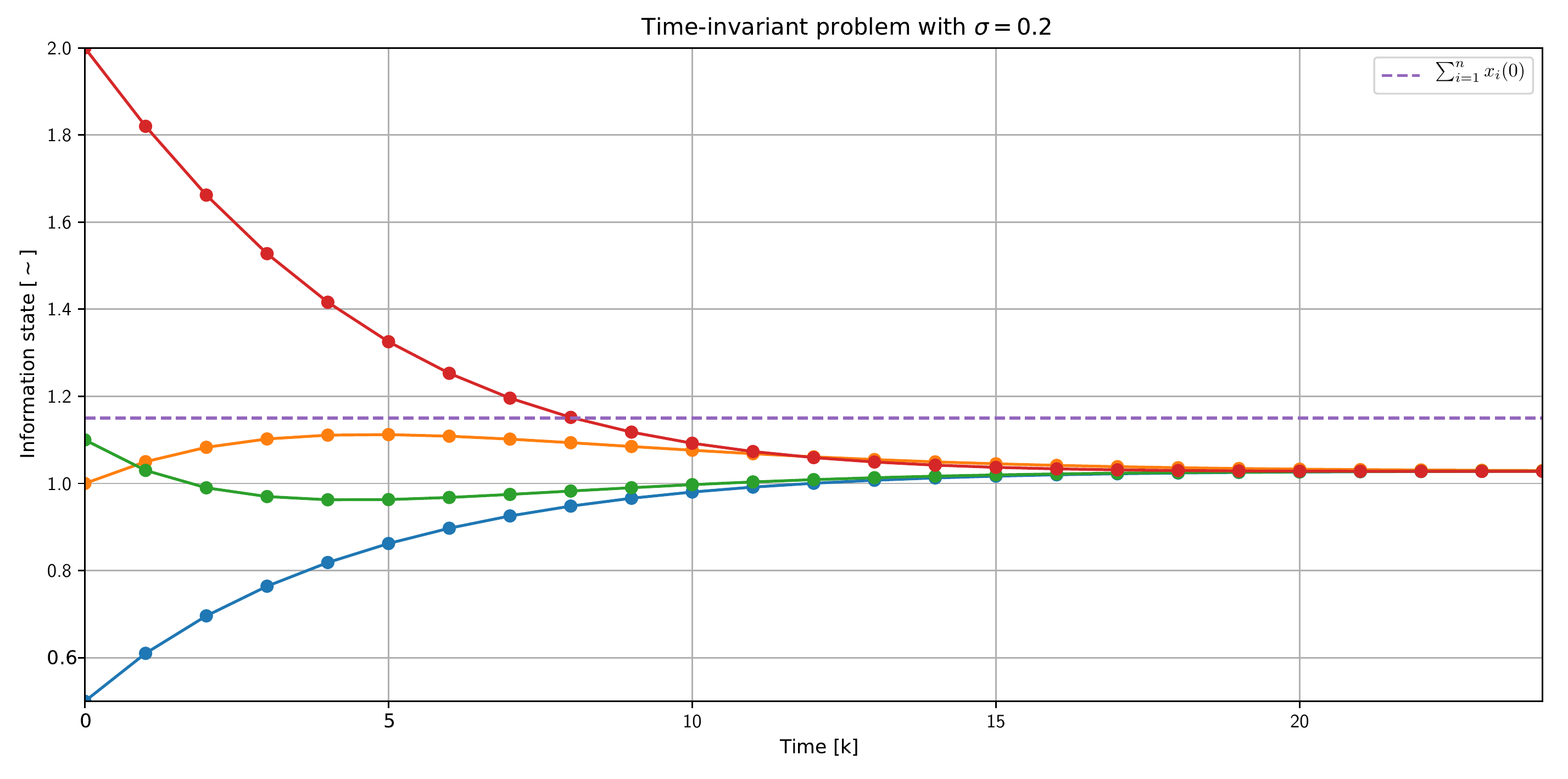}
			\caption{Time-invariant consensus problem with $\sigma_i=\sigma=0.2$.}
			\label{fig:consensus_ti_s2}
		\end{subfigure}
		\begin{subfigure}[b]{\columnwidth}
			\includegraphics[width=\columnwidth]{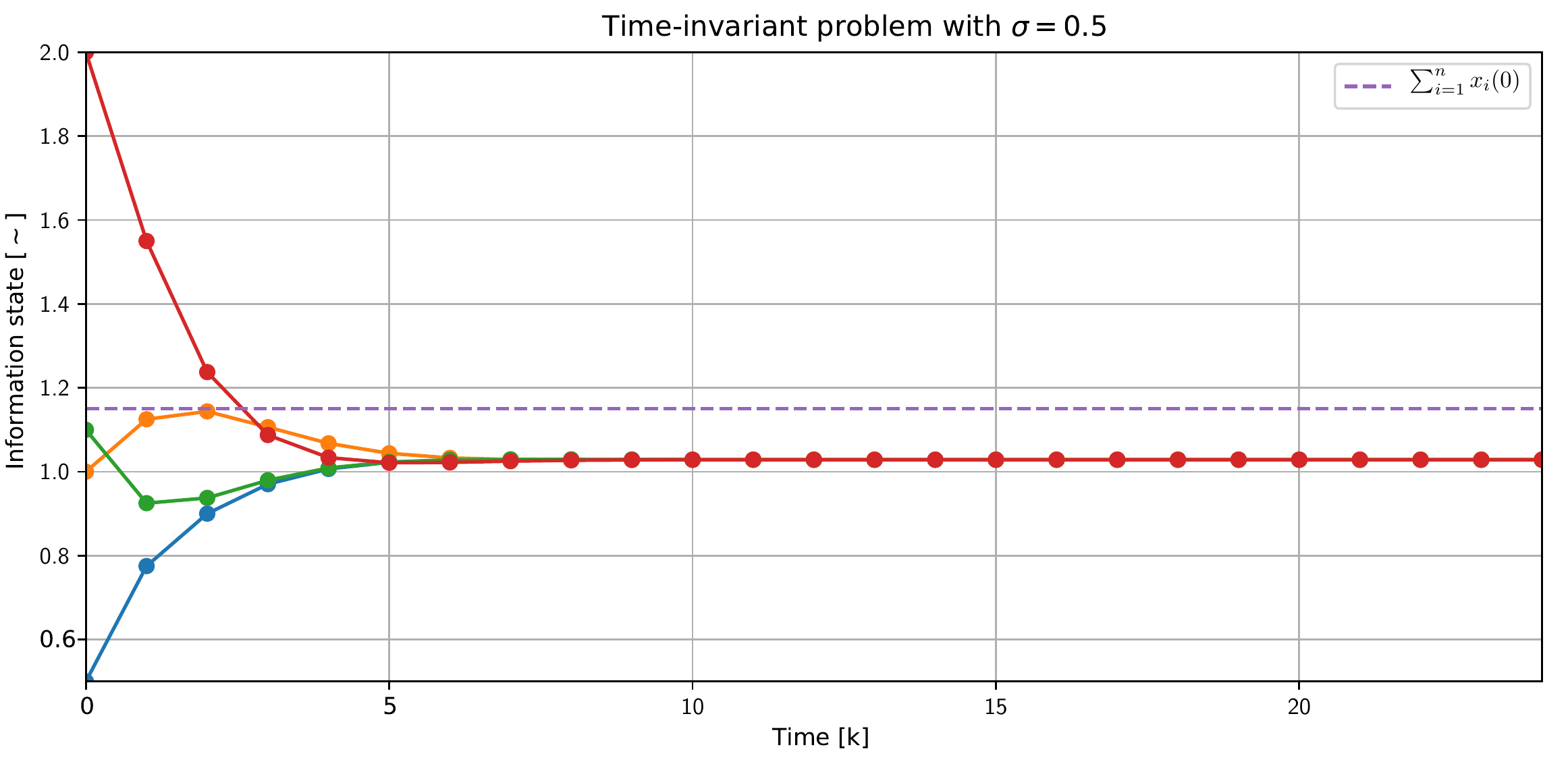}
			\caption{Time-invariant consensus problem with $\sigma_i=\sigma=0.5$.}
			\label{fig:consensus_ti_s5}
		\end{subfigure}	
		\caption{Consensus problem with constant channel coefficients.}
	\end{figure}

	Let us consider the balanced communication topology in \Cref{fig:comm_graph}. 	
	Let the initial information states $x_i(0)$ be randomly generated out of an uniform distribution between $0$ and $2\pi$, i.e. $x_i(0)\sim\mathcal{U}(0,2\pi)$, $\forall i\in\mathcal{N}$. First, consider the time-invariant system, where channel coefficients are distributed like $h_{ij}\sim\mathcal{U}(0,10)$, $\forall (j,i)\in\mathcal{A}$. The system in \Cref{fig:consensus_ti_s2}, with $\sigma_i=\sigma=0.2$, achieves weighted average consensus. 	
	Larger values of $\sigma$ give a higher rate of convergence, resulting in a faster system, as in \Cref{fig:consensus_ti_s5}, where the initial state vector and the realizations of channel coefficients are the same as in \Cref{fig:consensus_ti_s2}, but $\sigma=0.5$. 
	In this case, as already shown, ${x}^*$ is independent of $\sigma$, which affects therefore only the convergence rate.
	%By increasing $n$, the cardinality of $\mathcal{N}$, agents get closer to linear average agreement, as in \Cref{fig:consensus_n100_s6}, where $n=100$. 
%	\begin{figure}[htbp]
%		\begin{subfigure}[t]{\columnwidth}
%			\centering
%			\includegraphics[width=\columnwidth]{img/consensus_n100_s6}
%			\caption{$\sigma=0.6$}
%			\label{fig:consensus_n100_s6}
%		\end{subfigure}
%		\begin{subfigure}[t]{\columnwidth}
%			\centering
%			\includegraphics[width=\columnwidth]{img/consensus_n100_s9}
%			\caption{$\sigma=0.9$}
%			\label{fig:consensus_n100_s9}
%		\end{subfigure}
%		\caption{Time-invariant consensus problem with $|\mathcal{N}|=100$.}
%	\end{figure}
	\begin{figure}[htbp]
		\centering
		\includegraphics[width=\columnwidth]{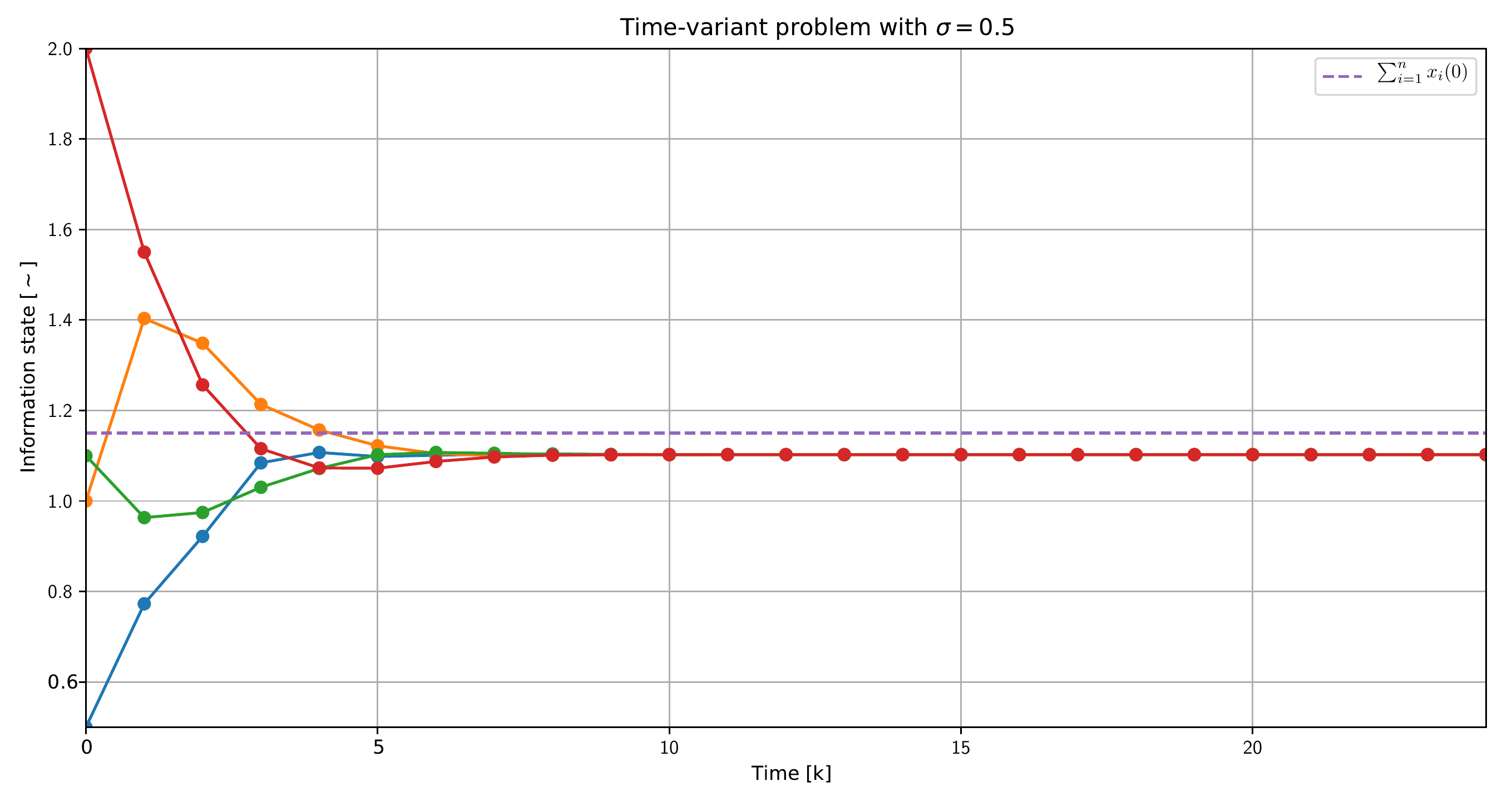}
		\caption{Time-variant system with $\sigma=0.5$.}
		\label{fig:consensus_tv_s5}
	\end{figure}
	
	We will now consider the time-variant case for $\sigma=0.5$ in \Cref{fig:consensus_tv_s5}. Channel coefficients are generated at each step under the assumption that they are independent and identically distributed. For the same $\sigma$, the convergence rate stays roughly the same as the one of the time-invariant case in \Cref{fig:consensus_ti_s5}. 	
	Also, as already proven, ${x}^*\in\mathcal{C}(\mathbf{x}(0))$. 
	
\begin{figure}[htbp]
		\centering
		\begin{subfigure}[b]{\columnwidth}
			\includegraphics[width=\columnwidth]{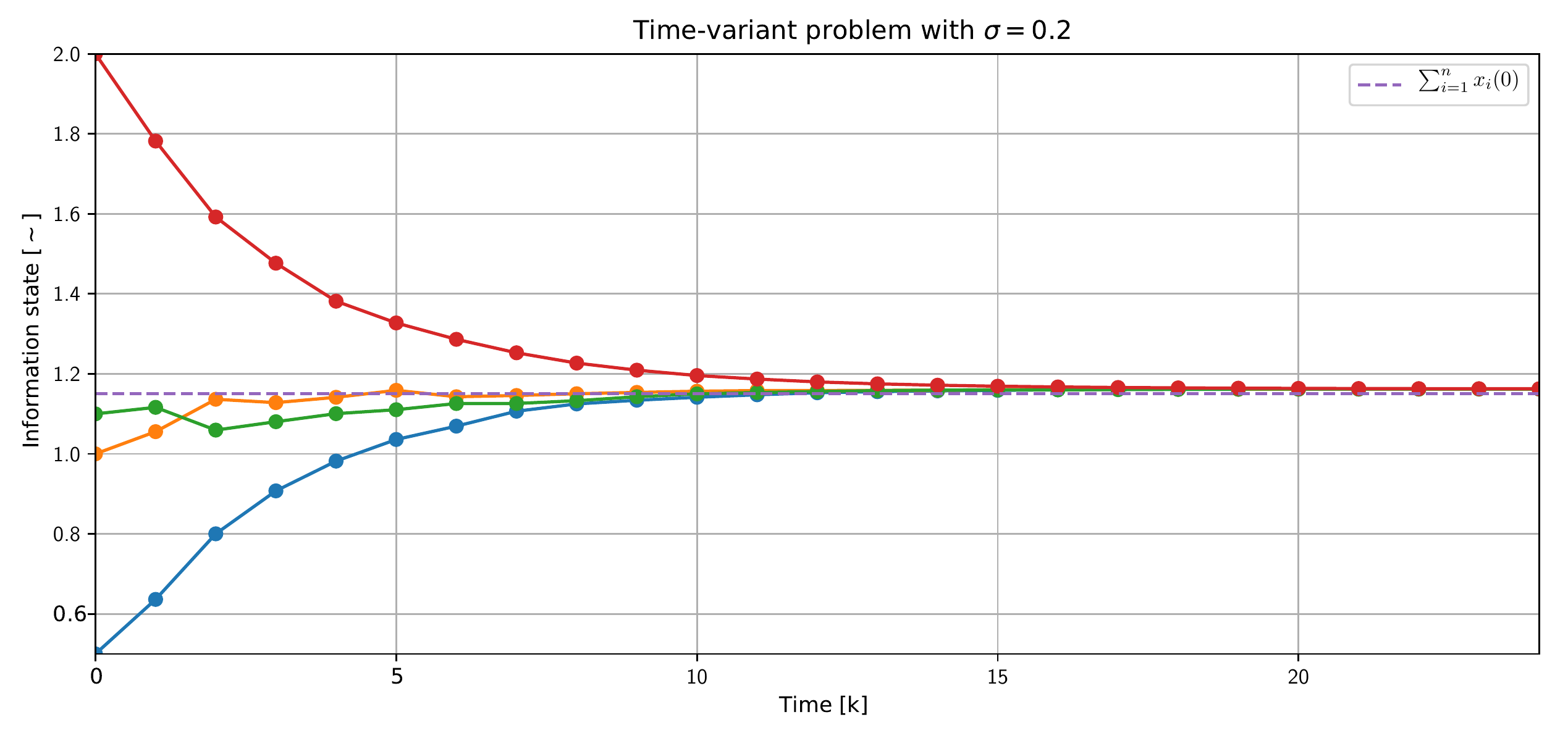}
			\caption{Time-variant system with $\sigma=0.2$.}
			\label{fig:consensus_tv_s2_fcn}
		\end{subfigure}
		\begin{subfigure}[b]{\columnwidth}
			\includegraphics[width=\columnwidth]{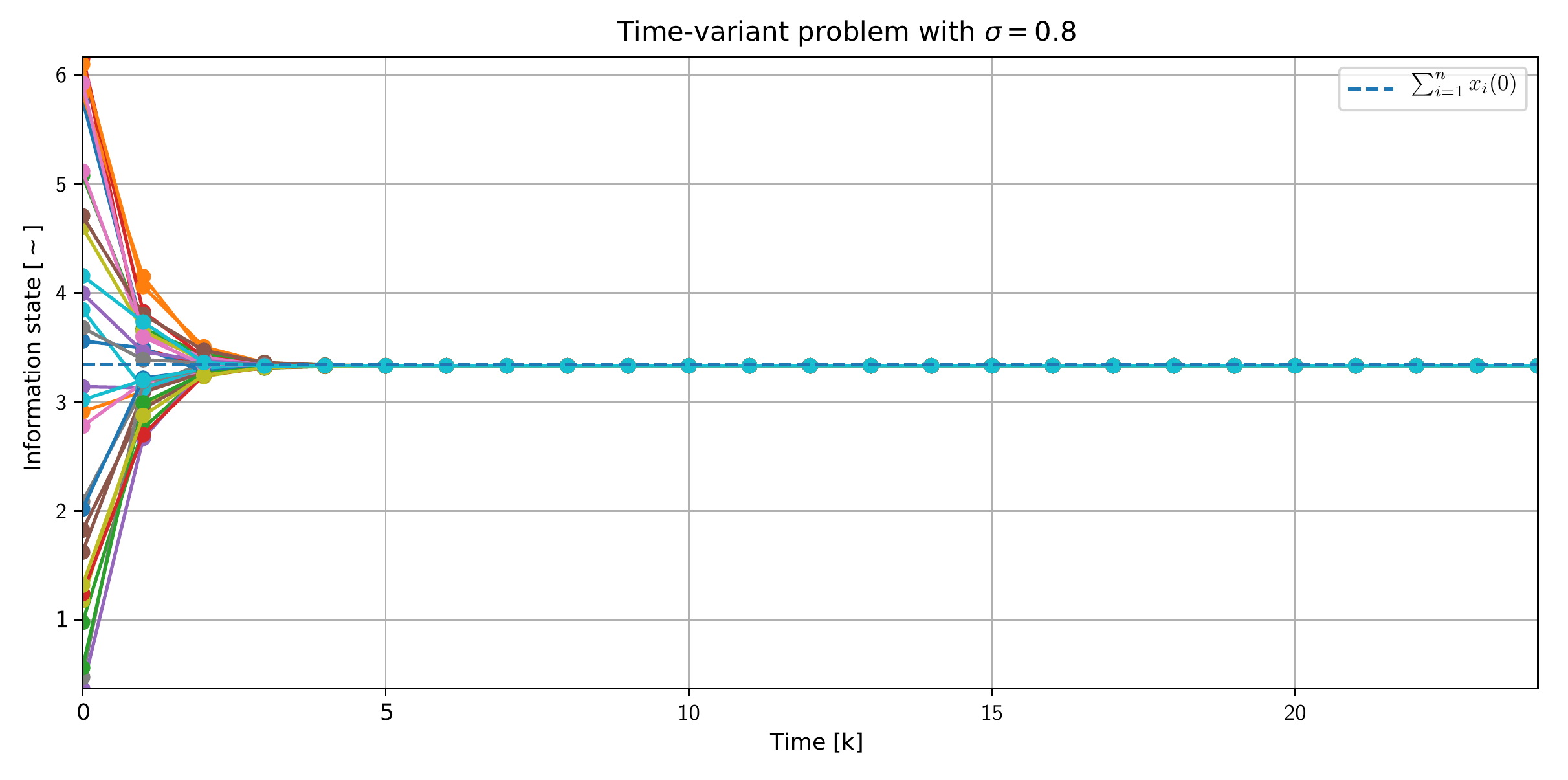}
			\caption{Time-variant system with $n=30$ and $\sigma=0.8$.}
			\label{fig:consensus_tv_s8_fcn_n30}
		\end{subfigure}
		\caption{Time-variant system over a fully connected communication topology.}
	\end{figure}

	According to \Cref{sec:influence}, for a fully connected network, in case of smaller $\sigma$ (more stubborn system), the rate of convergence is expected to be much slower, but the consensus value is closer to the linear average of initial information states, as in \Cref{fig:consensus_tv_s2_fcn}.
	
	By increasing the network size to $n=30$ and setting $\sigma$ to $0.8$, with a fully connected topology, agents achieve consensus (\Cref{fig:consensus_tv_s8_fcn_n30}), converging to an agreement value closer to the linear average consensus.
	
	As a conclusion, the proposed control offers robustness against the variation of positive channel coefficients (under \Cref{ass:ass1}); the system always achieves consensus. However, the agreement value is depending on the realizations of the channel coefficients. 
	
%	achieving weighted average consensus rather than linear average one is the cost to pay. Nevertheless, for systems involving a large number of agents or for time-variant systems with stubborn dynamics, the consensus value is closer to the linear average one.
	
%	Since it would be sufficient to study the eigenvalues of the dynamics matrix, a better analysis can be provided for the time-invariant system
%	\begin{equation}
%		\mathbf{x}(k+1)=D_n^\sigma \mathbf{x}(k),
%	\end{equation}
%	where, under the assumption that channel coefficients are the same through time, i.e. $h_{ij}(k_1)=h_{ij}(k_2)$, $\forall(i,j)\in\mathcal{N}\times {N}_i$, $\forall k_1,k_2\in \mathbb{N}$, the dynamics matrix $D_n^\sigma$ can be written as \cref{eq:matrixSigma} by omitting the time dependence; 
%%	\begin{equation}
%%		\label{eq:matrixSigmaInv}
%%		D_n^\sigma=
%%		\begin{bmatrix}
%%			(1-\sigma)	&\frac{\sigma h_{12}}{\sum_{j\in N_1}h_{1j}}	&\dots	&\frac{\sigma h_{1n}}{\sum_{j\in N_1}h_{1j}}\\
%%			\frac{\sigma h_{21}}{\sum_{j\in N_2}h_{2j}}	&(1-\sigma)	&\dots	&\frac{\sigma h_{2n}}{\sum_{j\in N_2}h_{2j}}\\
%%			\dots &\dots &\dots &\dots\\
%%			\frac{\sigma h_{n1}}{\sum_{j\in N_n}h_{nj}}	&\frac{\sigma h_{n2}}{\sum_{j\in N_n}h_{nj}}	&\dots	&(1-\sigma)
%%		\end{bmatrix}.
%%	\end{equation}
%	it is positive Perron matrix of an unbalanced graph for a suitable $\epsilon$. 
	%An analysis of the concept of algebraic connectivity for unbalanced graphs will be presented later on; a study of the impact that both $\sigma$ and fading coefficients have on convergence rate and $x^*$ will be presented.
	
	\section{Conclusion}
	\label{sec:conclusion}
	In this paper, we investigated a consensus scheme that exploits the superposition property of wireless communication by relying on broadcast of information. In particular, we took the case of unknown (time-variant and time-invariant) channel coefficients into account. In both cases, the resulting consensus is a weighted average one. We introduced a tuning parameter that, in the time-invariant case, influences the convergence rate, but not the asymptotic consensus. It was also shown, that this tuning parameter affects both convergence rate and value in the time-varying case. Finally, for systems with a large number of agents, the resulting consensus value will be close to the linear average value.
	
%	In this paper, we proposed a decentralized control strategy that achieves average consensus over wireless networks in the presence of channel coefficients. The cost to pay is having a weighted average consensus, rather than a desired linear average one. For the time-invariant problem, it is possible to properly tune some design parameters ($\sigma_i$), such that only convergence rate is affected and not the asymptotic consensus. Conversely, it was showed that tuning $\sigma$ in the time-variant case affects both the convergence rate and the consensus value. It is possible, then, to get closer to the ideal case of linear average consensus, by increasing the number of agents or by keeping more stubborn dynamics (this only for the time-variant case). 
	%Moreover, convergence rate for non-symmetric and non-double-stochastic Perron matrices has been linked to the algebraic connectivity of an associated undirected graph.
	
%	Future development will consider general \textit{f-consensus} problems and the impact that channel coefficients have in this context.
	
	\bibliographystyle{IEEEtran}
	\bibliography{bibliography}

\begin{thebibliography}{10}
\providecommand{\url}[1]{#1}
\csname url@samestyle\endcsname
\providecommand{\newblock}{\relax}
\providecommand{\bibinfo}[2]{#2}
\providecommand{\BIBentrySTDinterwordspacing}{\spaceskip=0pt\relax}
\providecommand{\BIBentryALTinterwordstretchfactor}{4}
\providecommand{\BIBentryALTinterwordspacing}{\spaceskip=\fontdimen2\font plus
\BIBentryALTinterwordstretchfactor\fontdimen3\font minus
  \fontdimen4\font\relax}
\providecommand{\BIBforeignlanguage}[2]{{%
\expandafter\ifx\csname l@#1\endcsname\relax
\typeout{** WARNING: IEEEtran.bst: No hyphenation pattern has been}%
\typeout{** loaded for the language `#1'. Using the pattern for}%
\typeout{** the default language instead.}%
\else
\language=\csname l@#1\endcsname
\fi
#2}}
\providecommand{\BIBdecl}{\relax}
\BIBdecl

\bibitem{degroot1974reaching}
M.~H. DeGroot, ``Reaching a consensus,'' \emph{Journal of the American
  Statistical Association}, vol.~69, no. 345, pp. 118--121, 1974.

\bibitem{ren2007information}
W.~Ren, R.~W. Beard, and E.~M. Atkins, ``Information consensus in multivehicle
  cooperative control,'' \emph{IEEE Control Systems}, vol.~27, no.~2, pp.
  71--82, 2007.

\bibitem{olfati2007consensus}
R.~Olfati-Saber, J.~A. Fax, and R.~M. Murray, ``Consensus and cooperation in
  networked multi-agent systems,'' \emph{Proceedings of the IEEE}, vol.~95,
  no.~1, pp. 215--233, 2007.

\bibitem{martinez2005robust}
S.~Martinez, J.~Cortes, and F.~Bullo, ``On robust rendezvous for mobile
  autonomous agents,'' \emph{IFAC Proceedings Volumes}, vol.~38, no.~1, pp.
  115--120, 2005.

\bibitem{fax2004information}
J.~A. Fax and R.~M. Murray, ``Information flow and cooperative control of
  vehicle formations,'' \emph{IEEE transactions on automatic control}, vol.~49,
  no.~9, pp. 1465--1476, 2004.

\bibitem{olfati2006flocking}
R.~Olfati-Saber, ``Flocking for multi-agent dynamic systems: Algorithms and
  theory,'' \emph{IEEE Transactions on automatic control}, vol.~51, no.~3, pp.
  401--420, 2006.

\bibitem{stanczak2009fundamentals}
S.~Stanczak, M.~Wiczanowski, and H.~Boche, \emph{Fundamentals of resource
  allocation in wireless networks: theory and algorithms}.\hskip 1em plus 0.5em
  minus 0.4em\relax Springer Science \& Business Media, 2009, vol.~3.

\bibitem{goldenbaum2012nomographic}
M.~Goldenbaum, H.~Boche, and S.~Sta{\'n}czak, ``Nomographic gossiping for
  f-consensus,'' in \emph{Modeling and Optimization in Mobile, Ad Hoc and
  Wireless Networks (WiOpt), 2012 10th International Symposium on}.\hskip 1em
  plus 0.5em minus 0.4em\relax IEEE, 2012, pp. 130--137.

\bibitem{buck1979approximate}
R.~C. Buck, ``Approximate complexity and functional representation,''
  \emph{Journal of Mathematical Analysis and Applications}, vol.~70, no.~1, pp.
  280--298, 1979.

\bibitem{zheng2012fast}
M.~Zheng, M.~Goldenbaum, S.~Sta{\'n}czak, and H.~Yu, ``Fast average consensus
  in clustered wireless sensor networks by superposition gossiping,'' in
  \emph{Wireless Communications and Networking Conference (WCNC), 2012
  IEEE}.\hskip 1em plus 0.5em minus 0.4em\relax IEEE, 2012, pp. 1982--1987.

\bibitem{ren2008distributed}
W.~Ren and R.~W. Beard, \emph{Distributed consensus in multi-vehicle
  cooperative control}.\hskip 1em plus 0.5em minus 0.4em\relax Springer, 2008.

\bibitem{moreau2005stability}
L.~Moreau, ``Stability of multiagent systems with time-dependent communication
  links,'' \emph{IEEE Transactions on automatic control}, vol.~50, no.~2, pp.
  169--182, 2005.

\bibitem{seneta2006non}
E.~Seneta, \emph{Non-negative matrices and Markov chains}.\hskip 1em plus 0.5em
  minus 0.4em\relax Springer Science \& Business Media, 2006.

\bibitem{wolfowitz1963products}
J.~Wolfowitz, ``Products of indecomposable, aperiodic, stochastic matrices,''
  \emph{Proceedings of the American Mathematical Society}, vol.~14, no.~5, pp.
  733--737, 1963.

\bibitem{sun2008average}
Y.~G. Sun, L.~Wang, and G.~Xie, ``Average consensus in networks of dynamic
  agents with switching topologies and multiple time-varying delays,''
  \emph{Systems \& Control Letters}, vol.~57, no.~2, pp. 175--183, 2008.

\bibitem{olfati2004consensus}
R.~Olfati-Saber and R.~M. Murray, ``Consensus problems in networks of agents
  with switching topology and time-delays,'' \emph{IEEE Transactions on
  automatic control}, vol.~49, no.~9, pp. 1520--1533, 2004.

\bibitem{goldenbaum2013harnessing}
M.~Goldenbaum, H.~Boche, and S.~Sta{\'n}czak, ``Harnessing interference for
  analog function computation in wireless sensor networks,'' \emph{IEEE
  Transactions on Signal Processing}, vol.~61, no.~20, pp. 4893--4906, 2013.

\end{thebibliography}
	
\end{document}